
\magnification\magstep1
\baselineskip=18pt
\rightline{OKHEP-95-05}
\rightline{Imperial/TP/94--95/32}
\medskip
\centerline{\bf Julian Schwinger:}
\centerline{\bf Source Theory and the UCLA Years---}
\centerline{\bf From Magnetic Charge to the Casimir Effect\footnote*
{\rm Invited Talk at Joint APS/AAPT Meeting, Washington, April 1995}}
\bigskip
\centerline{Kimball A. Milton}
\centerline{\sl Imperial College and the University of Oklahoma}
\bigskip

{\narrower\bigskip
ABSTRACT: Julian Schwinger began the construction of Source Theory in
1966
in response to the then apparent failure of quantum field theory
to describe strong interactions, the physical remoteness of
renormalization, and the utility of effective actions
in describing chiral dynamics.  This development did not meet with
wide acceptance, and in part for this reason Julian left Harvard
for UCLA in 1971.  This nonacceptance was quite understandable,
given the revolution in gauge theories that was then unfolding,
a revolution, of course, for which he had laid much of the
groundwork.
Acceptance of his ideas was further impeded by his rejection
of the quark model of hadrons and of QCD.
I will argue, however, that the source theory development was
not really so abrupt a break with the past as Julian may have
implied,
for the ideas and techniques in large measure were present in his
work
at least as early as 1951. Those techniques and ideas are still of
fundamental importance to theoretical physics, so much so that
the designation ``source theory'' has become superfluous.
Julian did a great deal of innovative physics during the last 30
years
of his life, and I will touch on some of the major themes, including
magnetic charge, chiral dynamics, radiation theory, Thomas-Fermi
models,
theory of measurement, and the Casimir effect, as well as various
forays
into phenomenology. The impact of much of this work is not yet
apparent.
\bigskip}


\vfill
\eject

Julian Schwinger, who died rather suddenly last July, was arguably
the most significant theoretical physicist since Dirac.
It is with great sadness that I will attempt to
summarize the second half of his  career, from 1966 until his
death in 1994, a time during which he published nearly 100 papers
on a great range of topics.  His passing was especially difficult
for me because at the time I,
ignorant of his illness, was on my way to work
with him on sonoluminescence; instead, when I arrived at UCLA
I spoke at the private memorial his wife Clarice held at their home,
and updated his complete publication list, a task I had commenced
seventeen years previously  in connection with his 60th birthday
festivities.\footnote{$^1$}{M. Flato, C. Fronsdal, and K. A. Milton,
{\it Selected Papers (1937--1976) of
Julian Schwinger} (Reidel,
Dordrecht, 1979). The numbers in this article enclosed in square
brackets refer to the list of Julian's papers I had compiled in
that reference, and updated in August of 1994. The updated list is
attached.}

\medskip
{\noindent{\bf Birth of Source Theory}}
\smallskip
I begin the tale with ``magnetic charge'' because his last
``operator''
field theory papers [129,130,133,134],
published in 1966,
and the first ``new'' papers after source theory was established
[147,150],
published in 1968--69, were devoted to that subject.
But probably the appropriate starting point is, in fact, his Nobel
lecture,  delivered on December 11, 1965 [132]. He ends the lecture
with
a discussion of phenomenological relativistic quantum field theory,
and states that ``One has still to appreciate the precise rules
of phenomenological relativistic field theory, \dots, given that
the strong fundamental interactions have operated to compose
the various physical particles.''  Is this not a prefigurement of
his attempt to create a source-theory revolution six months later?

It surely was the difficulty of incorporating strong interactions
into
field theory that led to ``Particles and Sources,'' received
by the Physical Review in July 1966 [135], a recording of lectures
Julian
gave in Tokyo that summer.  Particle phenomenology is primary, and
I personally note with relish that he cites my Oklahoma colleague
George Kalbfleisch in the second sentence of the introduction for
the discovery of the $\eta'$ meson.  This paper already included
particles
of all spins through the use of multispinors. The following year
there was an explosion of partial (chiral) symmetry papers
[137--41,143--45].  I believe that it was, in fact, his attempt
to put current algebra in effective Lagrangian language, together
with
Weinberg, which was the immediate impetus to the source-theory
development.  These papers were quite important at the time.

\medskip
\noindent{\bf What is Source Theory?}
\smallskip
Although Julian had invented the notion of a source at least as early
as 1951, it was only in 1966 that he realized that he could base the
whole machinery of particle physics on the abstraction of
particle-creation and annihilation acts.  One can define a free
action, say for a photon, in terms of propagation of virtual photons
between photon sources, conserved in order to remove the scalar
degree
of freedom.  But a virtual photon can in turn act as a pair of
electron-positron sources, through a ``primitive interaction''
between
electrons and photons, essentially embodied in the conserved Dirac
current.  So this multiparticle exchange gives rise to quantum
corrections to the photon propagator, to vacuum polarization, and so
on.  All this without any reference to renormalization or
``high-energy
speculations.''

In its ``purest'' or at least original form, such source theory ideas
were used to generate perturbative amplitudes in ``causal'' form;
that is, in which real particles were exchanged between virtual
sources separated in time.  From this one could deduce immediately
(``space-time extrapolation'') the full amplitude in spectral form,
that is, in what most people would refer to as a ``dispersion
relation.''  Such a direct
generation of amplitudes was extremely powerful, and often allowed
a completely finite calculation to be carried out.  An impressive
example is our calculation of the 4th-order Compton-scattering
helicity amplitudes directly in double-spectral
form.\footnote{$^2$}{K. A. Milton, L. L. DeRaad, Jr., and W.-y.\
Tsai,
Phys.\ Rev.\ D {\bf 6}, 1411 (1972)}
Noncausal methods, more reminiscent of usual Feynman diagram
techniques, but significantly different in spirit, were also
developed, and there we showed the power of the technique by some
very simple pen-and-ink calculations of 6th-order processes
contributing to the electron's magnetic moment.\footnote{$^3$}{K. A.
Milton, L. L. DeRaad, Jr., and W.-y.\ Tsai, Phys.\ Rev.\ D {\bf 9},
1809, 1814 (1974)}

So what is the legacy of the source-theory experience?  I think it is
more evolutionary than revolutionary.  New techniques were introduced
by Julian, principally in the causal formulation, that supplement
those
introduced in earlier decades, such as the proper-time technique
(which everyone uses nowadays), the quantum action principle
(particularly beloved by atomic physicists now), as so on, as
detailed by Lowell Brown. One commonality is the emphasis of the
power of differential, rather than integral techniques (``It
continues to surprise me that so many people seem to accept
this formal statement [the solution of the quantum action
principle as a path integral] as a satisfactory {\it starting\/}
point of a theory'' [160]).  Certainly in my own work that has been a
continuing theme,  even if the word source theory now occurs but
rarely.  I interpret the decision of the PACS indexers to remove
the ``source theory'' category not as a sign that source theory has
become irrelevant or redundant (in the British sense); but rather
that these useful techniques are part of the common language and
ammunition that theorists use to attack the most difficult problems
in physics.

Let us return to the history.
\medskip
\noindent{\bf Source Theory at Harvard}
\smallskip

In 1967 ``Source and Electrodynamics'' [142] was published, which
put QED into the new framework.  The following year, Julian treated
gravitons, and he gave {\it his\/}
 demonstration that full general relativity
is essentially a consequence of assuming that the mediator of the
gravitational force  is a massless helicity-2 particle
[146,162,163,177].
It was roughly at this point that I entered the picture, when,
as a second-year student, all fear and trembling, I asked Julian
if I could work for him. (But I was well prepared, bringing
a good knowledge of Green's functions from the University of
Washington.)
 I told him I was also taking Sydney Coleman's
field theory lectures and Arthur Jaffe's constructive field theory
course,
but that was all right with Julian, in spite of his plea for the
mind not ``warped \dots past the elastic limit.'' (The quotation is
from the preface of [153].)
The first book treatment of source theory, based on the Brandeis
lectures, appeared in 1969 [149]; Julian presented me with a copy for
successfully passing my oral exam (which I recall as primarily an
argument between Julian and Paul Martin).  I also recall the
excitement
of his source theory treatment of magnetic charge [147], particularly
his speculative dyon model of matter which he published in Science
in 1969 [150].  (His philosophy here was summed up in his quotation
from Faraday: ``Nothing is too wonderful to be true, if it be
consistent
with the laws of nature, and in such things as these,
experiment is the best test of such consistency,'' which I would
later find emblazoned
on the walls of the old physics building at UCLA, Kinsey Hall.)

Three other books came out in as many years: {\it Discontinuity
in Waveguides} (1968) [148], based on Dave Saxon's notes recording
a small portion of his wartime radar work; {\it Quantum Kinematics
and Dynamics} (1970) [152],  an unfinished textbook on quantum
mechanics, and {\it Particles, Sources, and Fields, Vol.~1} (1970)
[153]. The latter was intended to be a comprehensive treatment of
source theory, based on the motto ``if you can't join 'em, beat
'em.''
Harold, the ``hypothetical alert reader of limitless dedication,''
makes his appearance, and unlike a real student, is allowed to
interrupt, particularly when he has ``an historical gleam in
his eye.''
Julian started writing the second volume of
 this book during a six-month sabbatical in Tokyo in 1970;
on his return, he announced to his twelve or so graduate students
that he was leaving Harvard in February 1971 for UCLA.  Although I
had only
begun my fourth year at Harvard, I didn't have long to worry, for
half an hour later he informed me, Lester DeRaad, Jr., and Wu-yang
Tsai that he had arranged with UCLA to bring us along as postdocs.
Little did I guess that my affiliation with UCLA would last a decade!

\medskip
{\noindent{\bf Source Theory at UCLA}}
\smallskip
Why did he leave Harvard in 1971?  Certainly, he perceived a chilly
reception for source theory at Harvard, and thought (more or less
erroneously) that UCLA would be more hospitable.  But, probably at
least as important was the fact he had been at Harvard for 25 years,
and felt the need of a change.  The sunny climes of Southern
California,
where he could and did swim and play tennis every day were an
enormous
attraction.  Although it was billed as a temporary move, it was
always
clear to me that it was to be a permanent change.
Appropriately, LA greeted his
arrival with a major earthquake.
He soon bought a beautiful home in Bel
Air, with magnificent views of the city and the
ocean.\footnote{$^4$}{He also took the opportunity to
correct the error in his license
plates discussed by Lowell Brown.  Since California required at least
one letter in vanity plates (unfortunately not Greek), he chose
brevity and universality: {\sl A137Z}.}
  One thing Julian did not anticipate:
the caliber of graduate students at UCLA was far inferior to what he
was used to at Harvard.  Consequently, after more than
 70 Ph.D.'s at Harvard, I believe
only three  ever finished at UCLA (only a few more started).
(I can only recall Luis Urrutia, Walter Wilcox, and
 Greg Wilensky.)

Of course, also in 1971 gauge theories took off again, which doomed
general reception  to source theory.  Julian was very much aware of
what was going on, and proposed his own $U(2)$ version of the
``standard
model'' in 1972 [155], phenomenologically acceptable in those days.
(Shelly Glashow has already reminded us of his fundamental work in
making
the electroweak synthesis possible.)
[We self-styled ``sourcerer's apprentices'' contributed several
papers to the
development of the electroweak theory.]  For the next two years he
worked
very hard on the second volume of {\it Particles, Sources, and
Fields}
(proofed scrupulously by us three), devoted to electrodynamics, which
came out in 1973 [158].  Also in 1973 was the rebirth of
strong-field electrodynamics, with the publication of ``Classical
Radiation of Accelerated Electrons.\  II. A Quantum Viewpoint''
[156],
the first paper in which series having been published in 1949 [56].
(This illustrates the continuity of Julian's work, a subject to which
I will return.)  This led to a series of papers with Tsai
[159,176,186],
the last of which harkens back to a 1954 paper on the quantum
corrections to synchrotron radiation [78].
What Julian viewed as a prediction of $J/\psi$, in the form of a
proposal of an alternative mechanism for avoiding
strangeness-changing
neutral currents, appeared in the same year
[157], which, after the November revolution,
 was followed a series of related phenomenological forays
on the $\psi$ particles
[166,169--71,173].  In 1974 he wrote two papers on ``renormalization
group without renormalization group''
[164--65].\footnote{$^5$}{Sometime around this point
Clarice introduced me to
the lovely and talented Margarita Ba\~nos, daughter of fellow
physicist and Radiation Lab colleague Alfredo Ba\~nos.  We were
married three years later.}

With some very impressive work on electrodynamics (including methods
harking
back to his 1951 ``Gauge Invariance and Vacuum Polarization'' paper
[64]
and other classic papers,  an independent calculation of the
4th-order contribution to the electron's magnetic moment, and
 a revisiting of the axial-vector anomaly which he had
discovered in [64]) constituting
the first half of the third volume of PSF, he abandoned work on the
book
at the point where he had to face up to strong interactions.  (The
uncompleted third volume eventually came out in 1989 when
Addison-Wesley
repackaged the whole set [211].)  However, he was not about
to abandon high-energy physics, for in 1974 Julian continued his
iconoclastic interpretation of phenomenology with an alternative
viewpoint
of deep-inelastic scattering based on double spectral forms
(the precursor was the Deser-Gilbert-Sudarshan
representation\footnote{$^6$}{S. Deser, W. Gilbert, and E. C. G.
Sudarshan, Phys.\ Rev.\ {\bf 115}, 731 (1959)}), work
which continued until 1977
[167,178,179,179a,181--83], starting from  the valid premise
 that scaling does not necessitate point-like constituents.

Julian revisited magnetic charge in 1975 [172], just in time to hope
that
``the Price might be right'' (paraphrased from {\it Selected
Papers}).\footnote{$^7$}{Julian often had the television on while
doing physics.}
 A joint analysis of ``dyon-dyon scattering''
followed in 1976 [180].  He also became interested in the Casimir
effect
in 1975 [174], I think through conversations with Seth Putterman.
We wrote some joint papers on the Casimir effect in 1978, among
other things reconfirming Tim Boyer's surprising result on the sign
of the spherical effect [187--88]. In 1977 or '78  Julian invited
Stan
Deser to UCLA to give us some private lessons on supersymmetry;
although he submitted a paper on the multispinor basis of
supersymmetry
in 1978 [190], he kicked himself for not thinking of the idea first:
In his words, ``All right, wise guy!  Then why didn't you do it
first?''

During all these years he taught brilliant graduate and undergraduate
courses in field theory (source theory) and quantum mechanics,
lecturing for two hours a day, twice a week, followed by lunch
with Bob Finkelstein and us.  At first we ate at various Chinese
restaurants, but then, as he became more diet conscious, at the
Chatam
in Westwood, where he always ordered rare roast beef.  Tennis with
Lester DeRaad was a regular part of his weekly regime.

I think it was in 1977 that Julian taught graduate electrodynamics,
in
a typically novel and very insightful way (including variational
principles,
of course, but especially noteworthy for the pre\"eminence of physics
over
mathematics),
and I suggested we turn the
notes into a textbook.  We completed a first draft (more properly,
version 1.5) of a manuscript, all
neatly typed by Gilda Reyes of UCLA, and signed a contract with W. H.
Freeman.  Unfortunately, about the time I left UCLA in 1979 Julian
decided
the manuscript did not sound enough like himself, and started
rewriting,
resulting in turgidity.  The project was abandoned in 1981.  However,
I taught electrodynamics last fall (scheduled before Julian's death),
and will do so again next year, so I have hope of reviving the book.

\medskip
\noindent{\bf The Last 15 Years}
\smallskip
In 1980, after teaching a quantum mechanics course
(a not-unusual sequence of events), Julian began a series
of papers on the Thomas-Fermi model of atoms [192--96,201--6].  He
soon
hired Berthold-Georg Englert replacing me as a postdoc to help with
the elaborate
calculations.  This endeavor lasted until 1985.\footnote{$^8$}{I
understand from conversations after my talk in Washington
that this work not only is regarded as important in its own right by
atomic physicists, but has led to some significant results in
mathematics.  A long series of substantial papers by C. Fefferman
and L. Seco has been devoted to proving his conjecture about the
$Z$ dependence of the ground state energy of large atoms [193],
starting with Bull.\ Am.\ Math.\ Soc.\ {\bf 23}, 525 (1990)
and continuing through Adv.\ Math.\ {\bf 111}, 88 (1995). }
In 1985 his popular book on relativity, {\it Einstein's Legacy},
appeared, based on a series of television programmes he presented for
the Open University in the UK some years earlier.
(Another legacy of those programmes was the robot who graced his
living room thereafter.)
 He wrote three ``Humpty
Dumpty'' measurement theory papers (dealing with spin coherence in a
Stern-Gerlach interferometer) in 1988 [208--10], in collaboration
with Marlan Scully and Englert.
Those who have taken his quantum mechanics courses know how central
the
Stern-Gerlach experiment was to his formulation of quantum mechanics.
He seemed to be spending a great deal of time on several book
projects, but to my knowledge, nothing was completed.
 He also wrote three very interesting
homages in the '80's: ``Two Shakers of Physics''
[200], the pun in the title referring to himself and Tomonaga,
``Hermann Weyl and Quantum Kinematics'' [208a] in which he
acknowledges
his debt to one of his ``gods,''  whose ways ``are mysterious,
inscrutable, and beyond the comprehension of ordinary mortals,''
and
``A Path to Electrodynamics'' [212], dedicated to Richard Feynman.
In 1989 began a series of papers on cold fusion [213--4,216--20],
about
which the less said, the better.  His last physics endeavor, as I
implied above, was the suggestion that the puzzling phenomenon
of sonoluminescence may be due to the
``dynamic Casimir effect'' [221--8]. (The last paper was submitted
on April 30, 1993.)
 Typically, he was unaware of some
of my own papers relevant to the subject, but, atypically, he  was
very explicitly seeking my collaboration in the last year of his life
(I talked to him at some length in December 1993, at the annual
Christmas party given by the Ba\~nos', which he and Clarice always
attended, and at a subsequent lunch). He felt that ``carrying out
that program is---as one television advertiser puts it---job one''
[229].
 Jack Ng and I are indeed
in the process of doing just that.

\medskip
\noindent{\bf Conclusions}
\smallskip
How do we place this portion of Julian's career in context? It seems
to me that a number of general conclusions may be drawn.

\item{1.} I would argue that source theory was not so abrupt a break
with the past as Julian presented it.  It becomes increasingly clear
as one reads PSF, or his general {\it oeuvre},
 that he returns to techniques he invented in the
'40's and '50's.
Examples are ``non-causal methods'' which can be found in his famous
1951 ``Gauge Invariance and Vacuum Polarization'' paper [64], strong
field
methods, which go back to his early work on synchrotron radiation
[56,78] (and also GIVP), and even the theory of sources, which he
introduced
also in 1951 [66].  He, of course, was aware of this continuity; but
he
felt the need to emphasize a rather complete break.  He saw a great
improvement in conceptual clarity, for when he did operator field
theory he carried around a great deal of baggage (which {\it
really\/}
is essential) which most people had dispensed with or ignored.
Source theory enabled {\it Julian\/} to dispense with the
``physical remoteness'' [153] of renormalization and confront the
physics
directly.  Undoubtedly, with hindsight, we can say that his later
work
would have had much greater impact if he had not drawn such an
exclusive distinction.
\item{2.} Of course, probably a bigger impediment to the reception
of his ideas was a change in the times.  Dispersion relations
had died before he mounted his attack, and field theory was reborn
with
the discovery by 't Hooft that gauge theories of weak and strong
interactions made sense.  He could accept the electroweak synthesis
(to which he had contributed so much), but not quarks and QCD.
The notion of ``particles'' which were not asymptotic states was
too distasteful. (Yet his idea of dyons was not so different---maybe
it was just the ``unmellisonant'' name [150].)
\item{3.} In many of his later projects, the first paper in the
series
was far and away the strongest.  He had a very useful idea in the
first
deep inelastic scattering paper [167], but thereafter the work
largely
reduced to fitting data with many parameters.  Although I am less
familiar  with that work, a similar characteristic is true of the
Thomas-Fermi papers (although here it is the first two papers that
stand out).
And in the ``dynamic Casimir effect'' work
there is enough in these many short papers for about one
substantial article; the essential calculations have yet to
be carried out (some of Julian's approximations are, I believe,
erroneous);
and the relevance to sonoluminescence remains to be established.
\item{4.} The last 30 years of his life were not Julian's strongest
scientifically.  Certainly not for lack of ability: He remained
an awesome calculator and a brilliant expositor of unconventional
and clever ideas.  But the times had changed, and Julian was no
longer
the molder of ideas for theoretical physics.  He is sometimes
criticized for venturing into phenomenology---but in fact his first,
and quite substantial, papers were phenomenological.  [The
unfortunate
distinction between theory and phenomenology (not one that Julian
ever made) is a product of the last decade or so.]  Much of his
criticism
of QCD is quite valid---the theory remains on very tenuous ground,
and is more of a parametrization than the first-principles theory it
pretends to be.  GUTs and strings he found outrageous not because
of their theoretical failings but because he, quite rightly,
found the notion of a desert between 1 TeV and the Planck scale
completely unbelievable---this was, after all, his reason for
inventing source theory, to separate high-energy speculations from
models of low-energy phenomena.
\item{5.} As footnote 8 illustrates, we should not underestimate
the power of his work to have a continuing impact.  We can
confidently
expect future surprises.  This may be true as well of the many papers
in the attached list to which I have not referred, because they do
not fit into a well-defined pigeonhole.  I can only urge the reader
to read his papers, for riches are contained therein.

Eight months before his death, Julian made his first appearance on
the
internet (and his final publication in any form) with his
July 1993 Nottingham
lecture, ``The Greening of Quantum Field Theory:  George and I''
(hep-ph/9310283) [229]. This lecture provides a remarkable
overview of Julian's work from his own perspective.
 I commend his final words to you:  Like
George Green, ``he is, in a manner of speaking, alive, well, and
living
among us.''

\bigskip
\centerline{Acknowledgements}
\medskip
I am grateful to the UK PPARC for a Senior Visiting Fellowship
and Imperial College for its hospitality.  I thank UCLA for its
hospitality during the period when I updated Julian's publication
list, and the US Department of Energy for partial financial support.
I dedicate this article
to Julian Schwinger, the most brilliant physicist I have known,
and one of my very dearest friends, to whom I owe so much.

\vfill
\eject

\baselineskip=18pt
\overfullrule=0pt

\centerline{\bf Publications of Julian Schwinger, 1976--1994}
\centerline{[This updates the publication list found in}
\centerline{{\it Selected Papers (1937--1976) of Julian Schwinger}}
\centerline{edited by M. Flato, C. Fronsdal, and K. A. Milton
(Reidel, Dordrecht, 1979)]}

\bigskip

\item{175.} Magnetic Charge, in {\it Gauge Theories and Modern Field
Theory}, eds. R. Arnowitt and P. Nath (MIT Press, Cambridge, Mass.,
1976), p. 337.

\item{176.} Classical and Quantum Theory of Synergic
Synchrotron-Cerenkov Radiation
(with W.-y.\ Tsai and T. Erber), Ann.\ Phys.\ (N.Y.) {\bf 96}, 303
(1976).

\item{177.} Gravitons and Photons:
The Methodological Unification of Source Theory,
	Gen.\ Rel.\ and Grav.\ {\bf 7}, 251 (1976).

\item{178.} Deep Inelastic Scattering of Leptons,
Proc.\ Natl.\ Acad.\ Sci.\ USA {\bf 73}, 3351 (1976).

\item{179.} Deep Inelastic Scattering of Charged Leptons,
Proc.\ Natl.\ Acad.\ Sci.\ USA {\bf 73}, 3816 (1976).

\item{179a.} Deep Inelastic Scattering of Polarized
Electrons---A Dissident View, Talk
	presented at Symposium on High Energy Physics with Polarized Beams
	and Targets, Argonne Nat. Lab., August 22--27, 1976. (New York,
1976)
	pp.\ 288--305.

\item{180.} Nonrelativistic Dyon-Dyon Scattering
(with K. A. Milton, W.-y.\ Tsai, L. L. DeRaad, Jr., and D. C. Clark),
Ann.\ Phys.\ (N.Y.) {\bf 101}, 451 (1976).

\item{181.} Adler's Sum Rule in Source Theory, Phys.\ Rev.\ D {\bf
15}, 910 (1977).

\item{182.} Deep Inelastic Neutrino Scattering and Pion-Nucleon Cross
Sections,
	Phys.\  Lett.\ {\bf 67B}, 89 (1977).

\item{183.} Deep Inelastic Sum Rules in Source Theory,
Nucl.\ Phys.\ {\bf B123}, 223 (1977).

\item{184.} The Majorana Formula, Trans. N. Y. Acad.\ Sci.\ {\bf 38},
170
(1977). (Rabi 	Festschrift).

\item{185.} Introduction and Selected Topics in Source Theory,
in {\it Proceedings of Recent
	Developments in Particle and Field Theory}, Tubingen 1977
	(Braunschweig, 1979), pp.\ 227--333.

\item{186.} New Approach to Quantum Correction in Synchrotron
Radiation (with W.-y. Tsai), Ann.\ Phys.\ (N.Y.) {\bf 110}, 63
(1978).

\item{187.} Casimir Effect in Dielectrics (with L. L. DeRaad, Jr.
and K. A. Milton) Ann.\ Phys.\ (N.Y.) {\bf 115}, 1 (1978).

\item{188.} Casimir Self-Stress on a Perfectly Conducting Spherical
Shell (with K. A. Milton and L. L. DeRaad, Jr.), Ann.\ Phys.\ (N.Y.)
{\bf 115}, 388 (1978).

\item{189.} Introduction to Source Theory, with Applications to High
Energy Physics, {\it Proceedings of the Seventh Particle Physics
Conference}, University of Hawaii Press, pp.\ 341--481 (1978).

\item{190.} Multispinor Basis of Fermi-Bose Transformation,
Ann.\ Phys.\ (N.Y.) {\bf 119}, 192  (1979).

\item{191.} Relativistic Comets, Kinam, {\bf 1}, 87 (1979).

\item{192.}Thomas-Fermi Model: The Leading Correction,
Phys.\ Rev.\ A {\bf 22}, 1827 (1980).

\item{193.} Thomas-Fermi Model: The Second Correction,
Phys.\ Rev.\ A {\bf 24}, 2353 (1981).

\item{194.} New Thomas-Fermi Theory: A Test (with L. DeRaad, Jr.),
Phys.\ Rev.\ A {\bf 25}, 2399 (1982).

\item{195.} Thomas-Fermi Revisited: The Outer Regions of the Atom
(with B.-G. Englert),	Phys.\ Rev.\ A {\bf 26}, 2322 (1982).

\item{196.} The Statistical Atom: A Study.  University of Miami,
P.A.M. Dirac Birthday	Volume, 1982.

\item{197.} Quantum Electrodynamics, J. Physique {\bf 43}, 409
(1982).

\item{198.} Electromagnetic Mass Revisited, Found.\ Physics {\bf 13},
373 (1983).

\item{199.} Renormalization Theory of Quantum Electrodynamics:
An Individual View, in {\it The Birth of Particle Physics},
Cambridge University Press, p. 329 (1983).

\item{200.} Two Shakers of Physics, in {\it The Birth of Particle
Physics}, Cambridge University	Press, p. 354 (1983).

\item{201.} Statistical Atom: Handling the Strongly Bound Electrons
(with B.-G. Englert),	Phys.\ Rev.\ A {\bf 29}, 2331 (1984).

\item{202.} Statistical Atom: Some Quantum Improvements
(with B.-G. Englert), Phys.\ Rev.\ A {\bf 29}, 2339 (1984).

\item{203.} New Statistical Atom: A Numerical Study
(with B.-G. Englert), Phys.\ Rev.\ A {\bf 29}, 2353 (1984).

\item{204.} Semiclassical Atom (with B.-G. Englert), Phys.\ Rev.\
A {\bf 32}, 26 (1985).

\item{205.} Linear Degeneracy in the Semiclassical Atom
(with B.-G. Englert), Phys.\ Rev.\ A {\bf 32}, 36 (1985).

\item{206.} Atomic-Binding-Energy Oscillations
(with B.-G. Englert), Phys.\ Rev.\ A {\bf 32},  47 (1985).

\item{207.} {\it Einstein's Legacy: The Unity of Space and Time},
Scientific American Library, Vol.\  16 (1985).

\item{208.} Is Spin Coherence Like Humpty Dumpty? I.
Simplified Treatment (with B.-G. Englert and M. O. Scully),
Found.\ Phys.\ {\bf 18}, 1045 (1988).

\item{208a.} Hermann Weyl and Quantum Kinematics, in
{\it Exact Sciences and Their Philosophical Foundations}, ed.\
W. Deppert, K. H\"ubner, A. Oberschelp, and V. Weidemann,
Verlag Peter Lang, Frankfurt, 1988, pp.~107--129.

\item{209.} Is Spin Coherence Like Humpty Dumpty? II.\
General Theory (with M. O. Scully and B.-G. Englert),
Z. Phys.\ D {\bf 10}, 135 (1988).

\item{210.} Spin Coherence and Humpty Dumpty.\ III.\
The Effects of Observation (with M. O. Scully and B.-G. Englert),
Phys. Rev. A {\bf 40}, 1775 (1989).

\item{211.} {\it Particles, Sources, and Fields\/} (3 volumes),
Addison-Wesley, Redwood City, CA (1989).

\item{212.} A Path to Quantum Electrodynamics, Physics Today,
February 1989.
(Reprinted in {\it Most of the Good Stuff: Memories of Richard
Feynman},
	ed.\ L. M. Brown and J. S. Rigden, AIP, New York, 1993, p. 59.)

\item{213.} Cold Fusion: A Hypothesis, Z. Nat.\ Forsch.\ A {\bf 45A},
756 (1990).

\item{214.} Nuclear Energy in an Atomic Lattice I, Z. Phys.\ D {\bf
15}, 221 (1990).

\item{215.} Anomalies in Quantum Field Theory, in {\it Superworld
III}, Proceedings of the 26th
	Course of the International School of Subnuclear Physics, Erice,
	Italy, 7--15 August 1988 (Plenum, New York, 1990).

\item{216.} Phonon Representations, Proc.\ Natl.\ Acad.\ Sci.\ USA
{\bf 87}, 6983 (1990).

\item{217.} Phonon Dynamics, Proc.\ Natl.\ Acad.\ Sci.\ USA
{\bf 87}, 8370 (1990).

\item{218.} Reflecting Slow Atoms from a Micromaser Field
(with B.-G. Englert),	Europhysics Lett.\ {\bf 14}, 25 (1991).

\item{219.} Nuclear Energy in an Atomic Lattice,
Prog.\ Theor.\ Phys.\ {\bf 85}, 711 (1991).

\item{220.} Phonon Green's Function,
Proc.\ Natl.\ Acad.\ Sci.\ USA {\bf 88}, 6537 (1991).

\item{221.} Casimir Effect in Source Theory II,
Lett.\ Math.\ Phys.\ {\bf 24}, 59 (1992).

\item{222.} Casimir Effect in Source Theory III,
Lett.\ Math.\ Phys.\ {\bf 24}, 227 (1992).

\item{223.} Casimir Energy for Dielectrics,
Proc.\ Natl.\ Acad.\ Sci.\ USA {\bf 89}, 4091 (1992).

\item{224.} Casimir Energy for Dielectrics: Spherical Geometry,
Proc.\ Natl.\ Acad.\ Sci.\ USA {\bf 89}, 11118 (1992).

\item{225.} Casimir Light: A Glimpse,
Proc.\ Natl.\ Acad.\ Sci.\ USA {\bf 90}, 958 (1993).

\item{226.} Casimir Light: The Source,
Proc.\ Natl.\ Acad.\ Sci.\ USA {\bf 90}, 2105 (1993).

\item{227.} Casimir Light: Photon Pairs,
Proc.\ Natl.\ Acad.\ Sci.\ USA {\bf 90}, 4505 (1993).

\item{228.} Casimir Light: Pieces of the Action,
Proc.\ Natl.\ Acad.\ Sci.\ USA {\bf 90} 7285 (1993).

\item{229.} The Greening of Quantum Field Theory: George and I,
Lecture at Nottingham, July 14, 1993 (hep-ph/9310283).

\bigskip

\centerline{Compiled by K. A. Milton, August 9, 1994.}

\bye